\documentclass[twoside]{dis08}
\usepackage[latin1]{inputenc}
\usepackage[dvips]{graphicx,epsfig,color}
\usepackage{wrapfig,rotating}
\usepackage{amssymb,amsmath,array}

\begin{document}
\title{Summary of the Electroweak and Beyond the Standard Model Working Group}

\author{Christopher Hays$^1$, Michael Kr\"amer$^2$, David M. South$^3$, and 
Alexander Filip \.Zarnecki$^4$
\vspace{.3cm}\\
1- University of Oxford, Department of Physics \\
Oxford OX1 3RH, United Kingdom 
\vspace{.1cm}\\
2- Institute for Theoretical Physics, RWTH Aachen University\\
D-52056 Aachen, Germany
\vspace{.1cm}\\
3- Technische Universit\"at Dortmund - Experimentelle Physik V\\
D-44221 Dortmund, Germany
\vspace{.1cm}\\
4- Institute of Experimental Physics, University of Warsaw \\
   Ho\.za 69, 00--681 Warszawa, Poland
}

\maketitle

\begin{abstract}
A wide array of deep--inelastic--scattering and hadron collider
experiments have tested the predictions of the electroweak theory and
measured its parameters, while also searching for new particles and
processes.
We summarise recent measurements and searches that probe the Standard
Model to unprecedented precision.
\end{abstract}

\section{Introduction}
\label{sec:intro}

The production and study of particles through deep inelastic
scattering is one of the main tools in high energy physics and has led
to a comprehensive understanding of the fundamental particles and
interactions up to the electroweak energy scale.
Successful running of the HERA $ep$ collider, $\sqrt{s} = 319$~GeV,
has recently come to an end, and selected results are already
available from the complete $1$~fb$^{-1}$ data set.
The Tevatron continues to collide protons and antiprotons at $\sqrt{s}
= 1.96$~TeV, with new analyses based on up to $3$~fb$^{-1}$ of data.
Later this year the LHC will ramp up proton beams to $\sqrt{s} =
10$~TeV, to be followed next year by $\sqrt{s} = 14$~TeV $pp$
collisions.
In addition, data from the completed NuTeV fixed--target
incident--neutrino experiment have been further analysed to test
several hypotheses for the source of the long--standing $\sin^2\theta_W$
discrepancy.

\section{Electroweak Measurements}
\label{sec:ewk}

The Lagrangian describing the electroweak SU(2)$_{\rm L}
\times$U(1)$_{\rm Y}$ symmetry, before symmetry breaking, is:
\begin{equation}
{\cal{L}}_{\rm EW} = {\cal{L}}_{\rm g} + {\cal{L}}_{\rm f} + {\cal{L}}_{\rm h} + {\cal{L}}_{\rm y}.
\nonumber
\end{equation}
\noindent
The four components are the gauge term (${\cal{L}}_{\rm g}$), the
fermion term (${\cal{L}}_{\rm f}$), the Higgs term (${\cal{L}}_{\rm
h}$), and the Yukawa coupling term (${\cal{L}}_{\rm y}$).
After the symmetry breaking via the Higgs mechanism, the Lagrangian
describing the masses and interactions of the physical particles can
be written as:
\begin{equation}
{\cal{L}}_{\rm EW} = {\cal{L}}_{\rm K} + {\cal{L}}_{\rm NC} + {\cal{L}}_{\rm CC} + {\cal{L}}_H + 
        {\cal{L}}_{HV} + {\cal{L}}_{WWV} + {\cal{L}}_{WWVV} + {\cal{L}}_{\rm Y}. 
\nonumber
\end{equation}
\noindent
The Lagrangian now includes a kinetic term (${\cal{L}}_{\rm K}$), a neutral
current term (${\cal{L}}_{\rm NC}$), a charged current term
(${\cal{L}}_{\rm CC}$), a Higgs self--coupling term (${\cal{L}}_{H}$), a
Higgs--gauge interaction term (${\cal{L}}_{HV}$), a trilinear gauge
interaction term (${\cal{L}}_{WWV}$), a quartic gauge interaction term
(${\cal{L}}_{WWVV}$), and a Yukawa term (${\cal{L}}_{\rm Y}$).
Only an electromagnetic U(1)$_{\rm EM}$ symmetry survives.


Ongoing and completed experiments continue to test many of the terms
in the broken electroweak Lagrangian, and the LHC experiments are
making final sensitivity estimates before first data.

\subsection{Trilinear Gauge Couplings}
\label{sec:tgc}

The Lagrangian for trilinear gauge couplings is:
\begin{eqnarray}
\label{eq:tgc}
{\cal{L}}_{WWV} &=& -ig[(W_{\mu\nu}W^{\mu} - W^{\mu}W_{\mu\nu})
        (A^{\nu}\sin\theta_W - Z^{\nu}\cos\theta_W) \nonumber\\[1mm] 
        && \qquad + W_{\nu}W_{\mu}(A^{\mu\nu}\sin\theta_W - Z^{\mu\nu}\cos\theta_W)].
\end{eqnarray}
\noindent
Recent Tevatron measurements of $W\gamma$, $WZ$, $Z\gamma$ and $ZZ$
production have probed the trilinear gauge vertices of the electroweak
theory.
%
%
$W\gamma$ production at the Tevatron occurs predominantly through
separate $W$ and $\gamma$ radiation from quarks in the $t$-- and
$u$--channels, with a smaller contribution from the $s$--channel
triple--gauge vertex.
A study of this vertex can be made through its interference with the
other channels.
The interference produces a dip in the $Q_l \times \Delta y$
distribution, where $Q_l$ is the charge of the lepton from the $W$
boson decay and $\Delta y$ is the rapidity difference between the
lepton and the photon.
This dip is known as a {\it radiation amplitude zero} and appears at
$Q_l \times \Delta \eta \approx -1/3$, where the pseudorapidity $\eta$
equals the rapidity in the massless limit.

\begin{wrapfigure}{r}{0.5\columnwidth}
\vskip -0.05in
\centerline{\includegraphics[width=0.45\columnwidth]{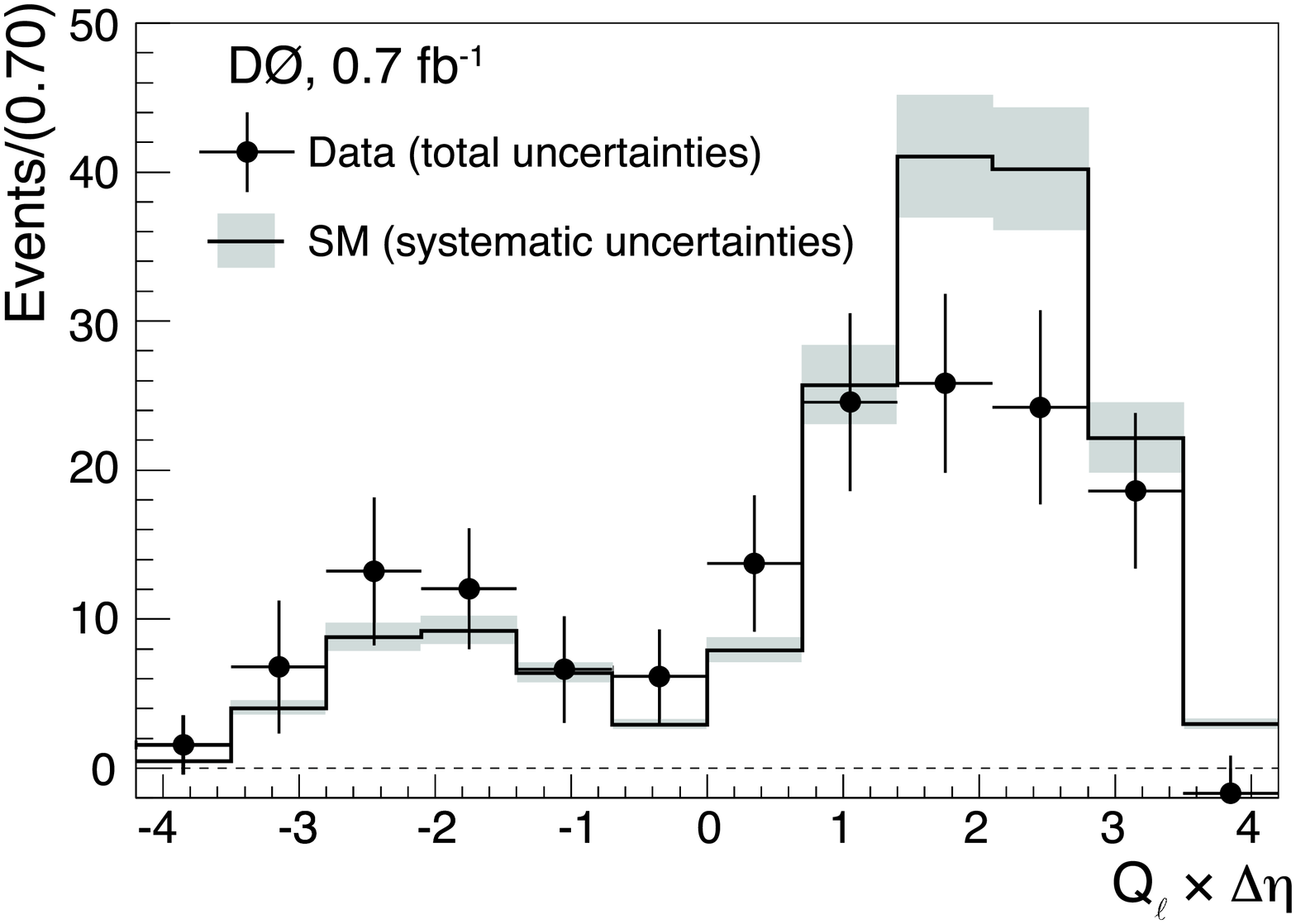}}
\vskip -0.1in
  \caption{The D\O\ charge--signed pseudorapidity difference between
  the charged lepton and the photon.  The dip at $Q_l \times \Delta
  \eta \approx -1/3$ is consistent with the interference due to the
  trilinear $WW\gamma$ coupling and has a significance of
  $2.6\sigma$. }
\label{fig:raz}
\end{wrapfigure}


D\O\ has probed the $l^{\pm}\nu\gamma$ final state for evidence of the
$s$--channel interference \cite{d0wg}.
A dip is observed consistent with the predicted radiation amplitude
zero, as shown in Figure~\ref{fig:raz}, and has a $2.6\sigma$
significance relative to a monotonic curve.
D\O\ has also searched for anomalous triple--gauge couplings that
modify the $W_{\nu}W_{\mu}A^{\mu\nu}$ term in Equation~\ref{eq:tgc} by
a scale factor $\kappa_{\gamma}/(1 + \hat{s}/\Lambda^2)^2$, and that
introduce a new term $ig\lambda_{\gamma}/[(1 + \hat{s}/\Lambda^2)^2
M_W^2] W_{\lambda\mu}W_{\nu}^{\mu}A^{\nu\lambda}$, for a cut--off
energy scale $\Lambda = 2$~TeV.
These terms enhance the production cross section for high--momentum
photons, and since no enhancement is observed, limits of $0.49 <
\kappa_{\gamma} < 1.51$ and $-0.12 < \lambda_{\gamma} < 0.13$ are
placed on these couplings.

\begin{wrapfigure}{!r}{0.5\columnwidth}
\centerline{\includegraphics[width=0.45\columnwidth]{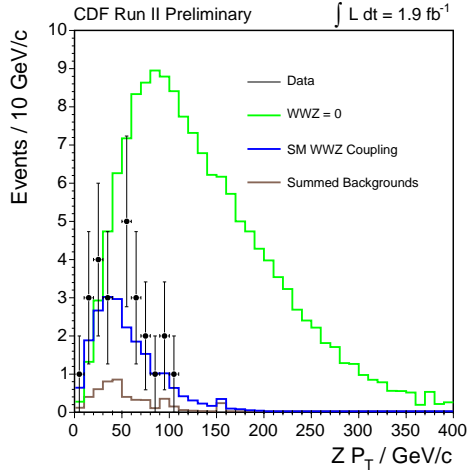}}
\vskip -0.1in
  \caption{The $P_{T}$ of the reconstructed $Z$ boson in $WZ$
  production at CDF.  The interference between diagrams with and
  without the $WWZ$ vertex significantly suppresses the cross section,
  and the data are consistent with this suppression. }
\label{fig:wz}
\end{wrapfigure}


The Tevatron is currently the only place to separately study the $WWZ$
and $WW\gamma$ vertices.
As with $W\gamma$ production, $WZ$ production at the Tevatron includes
both an $s$--channel diagram with a triple--gauge vertex and $t$-- and
$u$--channel diagrams where the bosons are radiated off the quarks.
Destructive interference is large in $WZ$ production, and the $WWZ$
vertex significantly reduces the cross section, as shown in
Figure~\ref{fig:wz}.
CDF has observed $WZ$ production at the $6\sigma$ level \cite{wz} in
the $l\nu l'l'$ final state, measuring a cross section
\begin{equation}
\sigma_{WZ} = 4.4^{+1.3}_{-1.0~{\rm stat}}~\pm~0.4_{\rm \; sys + lum}~{\rm pb}
\nonumber
\end{equation}

\noindent
that is consistent with the next--to--leading order (NLO) prediction
of $\sigma_{WZ} = 3.7 \pm 0.3$ pb.
Both CDF and D\O\ set limits (Table~\ref{tbl:wz}) on the anomalous
coupling parameters $\lambda_Z$ and $\Delta \kappa_Z = \kappa_Z - 1$,
which are the analogues to the $WW\gamma$ vertex parameters.
Limits are also set on $\Delta g_1^Z = g_1^Z - 1$, where $g_1^Z$ is a
scale factor for the $(W_{\mu\nu}W^{\mu} - W^{\mu}W_{\mu\nu})$ term.


The Standard Model (SM) does not have trilinear gauge vertices
involving only neutral particles.
The Tevatron experiments test this prediction by studying $Z\gamma$
and $ZZ$ production.
CDF and D\O\ select $Z\gamma$ events in the $ll\gamma$ final state,
requiring $E_T^{\gamma} > 7$ GeV and $\Delta R(l,\gamma) > 0.7$.
Using $778$ and $968$ candidates respectively, CDF and D\O\ measure
$Z\gamma$ production cross sections of
\begin{eqnarray*}
\sigma_{Z\gamma}^{\rm CDF}   & \!\!= & \!\!  4.6~\pm~0.4_{\rm \; stat + sys}~\pm~0.3_{\rm\;lum}~{\rm pb} \\
\sigma_{Z\gamma}^{\rm D\O\ } & \!\!= & \!\! 5.0~\pm~0.3_{\rm \; stat + sys}~\pm~0.3_{\rm\; lum}~{\rm pb,} 
\end{eqnarray*}
\noindent
consistent with the NLO SM prediction of $\sigma_{Z\gamma} = 4.5 \pm
0.4$~pb. Limits are set on anomalous couplings using the consistency
of data and prediction at high photon momentum.

\begin{wraptable}{!r}{0.5\columnwidth}
\vskip -0.2in
\centerline{\begin{tabular}{|c||c||c|}
\hline
Coupling        & CDF           & D\O\          \\ \hline
\hline
$\lambda_Z$     & $(-0.13, 0.14)$ & $(-0.17, 0.21)$ \\ \hline
$\Delta g_1^Z$  & $(-0.13, 0.23)$ & $(-0.14, 0.34)$ \\ \hline
$\Delta \kappa_Z$ & $(-0.76, 1.18)$ & $(-0.12, 0.29)$$^*$ \\
\hline
\end{tabular}}
\caption{Limits on anomalous $WWZ$ couplings for a cut--off scale of $\Lambda = 2$ TeV.  
$^*$D\O\ limits on $\Delta \kappa_Z$ are for the case $\Delta \kappa_Z = \Delta g_1^Z$.}
\label{tbl:wz}
\end{wraptable}

The rare process of $ZZ$ production is now beginning to appear at the
Tevatron.
CDF has three candidate events in the four--lepton final
state \cite{cdfzz}, and with a background of only $0.1 \pm 0.1$ this
constitutes a $4.2\sigma$ evidence.
Combining with $276$ candidate events in the $ll\nu\nu$ channel, where
a likelihood discriminant is used to separate the $14$ expected $ZZ$
events, gives a total significance of $4.4\sigma$.
The measured cross section of $\sigma_{ZZ} = 1.4^{+0.7}_{-0.6}~{\rm
pb}$ is consistent with the NLO prediction of $\sigma_{ZZ} = 1.4 \pm
0.1$ pb.
Recently D\O\ has claimed a $5.7\sigma$ observation of $ZZ$
production~\cite{d0zz}.

\subsection{Charged and Neutral Currents}
\label{sec:ccnc}

\begin{wrapfigure}{!r}{0.48\columnwidth}
\vskip -0.1in
\centerline{\includegraphics[width=0.48\columnwidth]{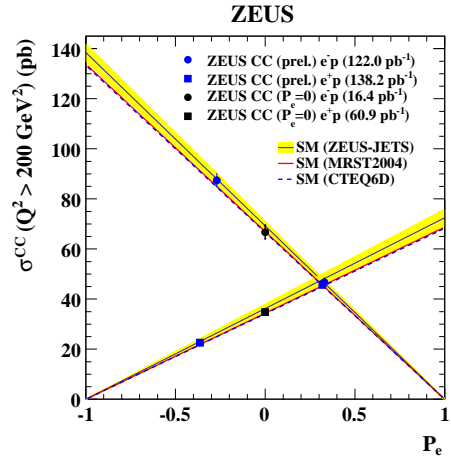}}
\vskip -0.1in
  \caption{The measured and theoretical charged current cross section
  in $e^{\pm}p$ collisions, as a function of lepton polarisation.  The
  ZEUS measurements are consistent with the $V - A$ structure of the
  weak interaction. }
\label{fig:ccxsec}
\end{wrapfigure}

The charged and neutral current contributions to the SM Lagrangian can
be expressed as:
\begin{eqnarray*}
{\cal{L}}_{\rm CC} & = & -\frac{g}{\sqrt{2}}\left[ \bar{u}_i \gamma^{\mu} \frac{1-\gamma^5}{2}M^{\rm CKM}_{ij}d_j
        + \bar{\nu}_i \gamma^{\mu} \frac{1-\gamma^5}{2} e_i \right]W_{\mu}^+ + {\rm h.c.}~~~{\rm and} \\
{\cal{L}}_{\rm NC} & = & eJ_{\mu}^{\rm em} A^{\mu} + \frac{g}{\cos\theta_W}(J_{\mu}^3 -
        \sin\theta_W J_{\mu}^{\rm em})Z^{\mu}.
\end{eqnarray*}

\noindent
The electromagnetic current $J_{\mu}^{\rm em}$ is vector--like ($V$),
coupling equally to left-- and right--handed helicity states.
The weak current $J^3_{\mu}$ is of a vector-minus--axial ($V-A$)
nature, coupling only to left--handed helicity states.
The CKM matrix $M^{\rm CKM}_{ij}$ determines the mixing between mass
and weak eigenstates of the quarks.

\begin{wrapfigure}{!r}{0.48\columnwidth}
\vskip -0.3in
\centerline{\includegraphics[width=0.48\columnwidth]{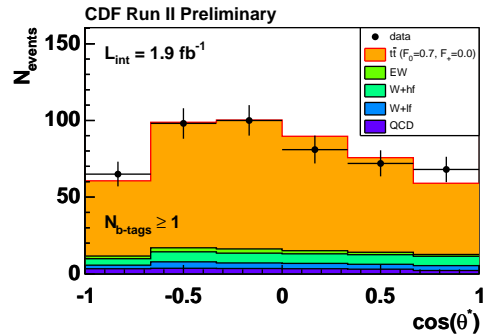}}
\vskip -0.1in
  \caption{The $\cos \theta^*$ distribution in $t\bar{t} \rightarrow
  WbWb \rightarrow l\nu b qq b$ events, where $\theta^*$ is the angle
  between the top quark and the charged lepton in the $W$ boson rest
  frame. }
\label{fig:whel}
\end{wrapfigure}

The electroweak couplings have no dependence on fermion generation.
This property has been tested with a measurement of the charged kaon
decay rate ratio $R_K = \Gamma(K^{\pm}\rightarrow
e^{\pm}\nu)/\Gamma(K^{\pm}\rightarrow \mu^{\pm}\nu)$ at the NA48 and
NA62 experiments at CERN.
The 2004 measurement of $R_K = 2.455 \pm 0.045_{\rm\; stat} \pm
0.041_{\rm\; sys} \times 10^{-5}$ is consistent with the SM prediction
of $R_K = 2.477 \times 10^{-5}$ and has a systematic uncertainty
dominated by the background prediction.
New methods to constrain the background using data are in progress,
promising a significant reduction in the uncertainty of this
measurement.
A recent review of precision SM tests with kaons can be found
in~\cite{wanke}.


Polarised electron beams in the $ep$ collisions at HERA allow a test of the 
$V-A$ structure of the weak coupling, represented by the $(1 - \gamma^5)/2$ 
factor in the Lagrangian.
The cross section of the charged current process 
$ep \rightarrow \nu X$ is a linear function of lepton polarisation, going to 
zero for right--handed electrons and left--handed positrons.
The ZEUS Collaboration has measured this cross section for
polarisations of $-0.27$ and $0.33$ for $e^- p$ collisions and $-0.36$
and $0.32$ for $e^+ p$ collisions, where the polarisation $P_e$ equals
one for right--handed lepton beams~\cite{zeusCC}.
Combining with unpolarised measurements, the observed cross section
dependence on polarisation is consistent with the SM, as shown in
Figure~\ref{fig:ccxsec}.

\begin{wrapfigure}{!r}{0.5\columnwidth}
\centerline{\includegraphics[width=0.45\columnwidth]{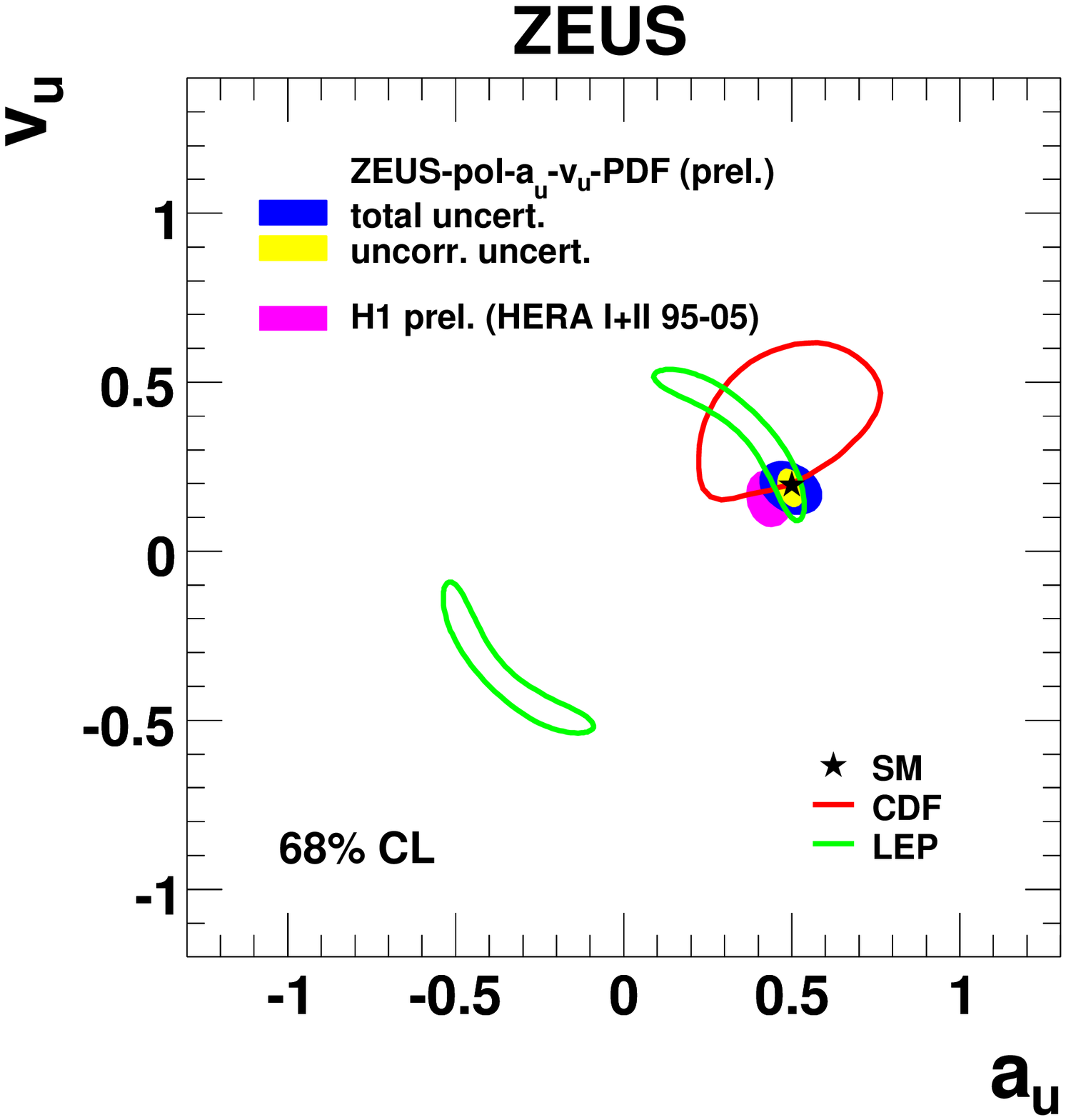}}
\centerline{\includegraphics[width=0.45\columnwidth]{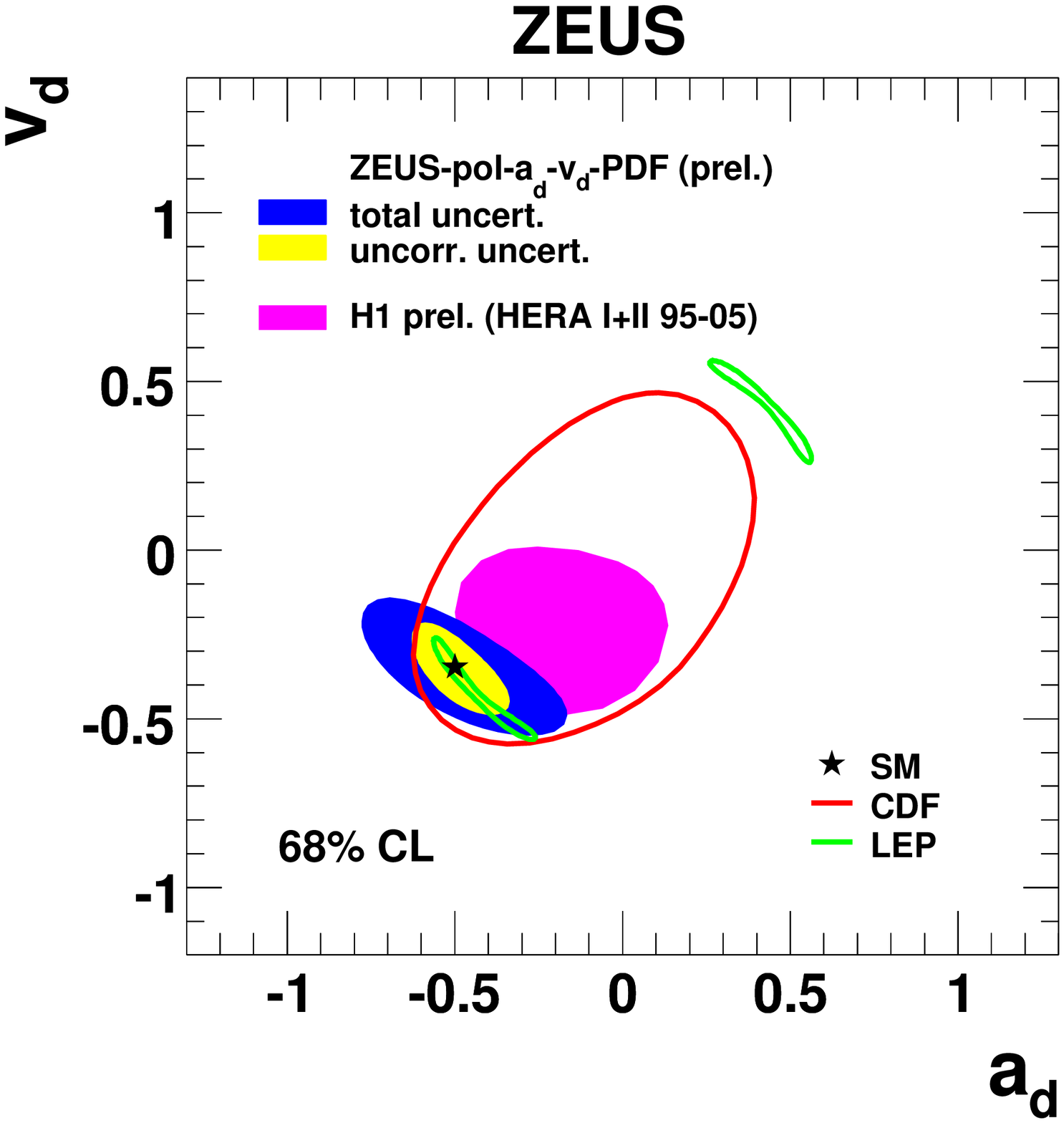}}
\vskip -0.1in
  \caption{The vector and axial neutral current couplings for up
  (top) and down (bottom) quarks from ZEUS, H1, CDF, and LEP. }
\label{fig:vqaq}
\end{wrapfigure}

Another test of the $V-A$ nature of charged current interactions
occurs through a measurement of the $W$ boson helicity in top quark
decays.
The helicity has a longitudinal component due to the $W$ boson mass
and a left--handed component due to the $W$ boson coupling.
CDF uses several techniques to measure the helicity in
$t\bar{t}\rightarrow WbWb \rightarrow l\nu b qq b$
events~\cite{cdfhelicity}.
One method uses the relative angle ($\theta^*$) between the top quark
and final--state charged lepton in the $W$ boson rest frame.  The
measured distribution of $\cos\theta^*$ is consistent with the SM
prediction of $70\%$ longitudinal helicity, as can be seen in
Figure~\ref{fig:whel}.
The most precise CDF and D\O\ measurements of the longitudinal
helicity of the $W$ boson in top decays are $0.64 \pm 0.08_{\rm\;
stat} \pm 0.07_{\rm\; sys}$ and $0.62 \pm 0.09_{\rm\; stat} \pm
0.05_{\rm\; sys}$, respectively.


The CKM matrix elements $M^{\rm CKM}_{ij}$ arise from off--diagonal
Higgs--quark Yukawa couplings in generation space before symmetry
breaking.
These couplings are not predicted by the SM and must be measured.
The element $V_{tb}$ can be determined with a measurement of single
top production at the Tevatron, since the cross section of the process
$p\bar{p} \rightarrow W \rightarrow tb$ includes a factor of
$|V_{tb}|^2$.
Both CDF and D\O\ \cite{d0tb} have obtained evidence for single top
production, with measured cross sections and $|V_{tb}|$ limits of
\begin{eqnarray*}
\sigma_{tb}^{\rm CDF} & = & 2.2~\pm~0.7~{\rm pb }~(|V_{tb}| > 0.66) \\
\sigma_{tb}^{\rm D\O\ } & = &  4.7~\pm~1.3~{\rm pb }~(|V_{tb}| > 0.68). 
\end{eqnarray*}
\noindent
Experiments at the LHC will also measure single top production in the
process $pp \rightarrow W \rightarrow tb$, and expect to constrain
$|V_{tb}|$ to $5\%$ with $10$~fb$^{-1}$ of data.


The fermion vector ($v_f$) and axial ($a_f$) couplings to the neutral
current are determined by the relative ${\rm SU(2)}_{\rm L}$ and ${\rm
U(1)}_{\rm Y}$ couplings $g$ and $g'$ respectively. 
These couplings were probed in electron--quark interactions at HERA,
through neutral current photon and $Z^0$ exchange in the $t$-- and
$u$--channels.
The experiments extract the vector ($v_q$) and axial ($a_q$) quark
couplings from a global fit to the neutral current and
charged current inclusive and jet cross sections.
The inclusion of charged current data constrains the parton
distribution functions, which are needed to extract the couplings.
Figure~\ref{fig:vqaq} shows the latest fits for $v_u$, $a_u$, $v_d$,
and $u_d$ from ZEUS~\cite{zeusEW} and H1~\cite{h1EW} global fits.


The weak mixing angle is defined in terms of the ${\rm SU(2)}_{\rm L}$
and ${\rm U(1)}_{\rm Y}$ couplings using the parameterisation $\tan
\theta_W = g'/g$.
By measuring neutral current and charged current cross sections in
neutrino--nucleon and antineutrino--nucleon scattering, $\sin^2
\theta_W$ can be extracted from the Paschos--Wolfenstein relation:
\begin{equation}
R = \frac{\sigma^{\nu}_{\rm NC} - \sigma^{\bar{\nu}}_{\rm NC}}{\sigma^{\nu}_{\rm CC} - \sigma^{\bar{\nu}}_{\rm CC}} = 
        \rho^2 \left(\frac{1}{2} - \sin^2\theta_W\right).
\nonumber
\end{equation}
\noindent
The NuTeV experiment measured $\sin^2\theta_W = 0.2277 \pm
0.0013_{\rm\; stat} \pm 0.0009_{\rm\; sys}$, which is nearly $3\sigma$
above the latest fit to electroweak data ($\sin^2\theta_W = 0.2231 \pm
0.0003$).
The discrepancy is dominantly in the neutrino cross sections, where
the ratio of neutral current to charged current cross sections
($R^{\nu}$) is nearly $3\sigma$ below expectation ($\Delta R^{\nu} =
-0.0034 \pm 0.0013$).
NuTeV has updated $R^{\nu}$, and the equivalent ratio for
antineutrino--nucleon scattering ($R^{\bar{\nu}}$), factoring in
several new results: a fit to its charm--decay data for a possible
strange--antistrange asymmetry in the nucleon; the latest
$K^+\rightarrow \pi^0 e^+ \nu$ ($K^+_{e3}$) branching ratio; and an
improved $d/u$ PDF uncertainty estimate.
The effects respectively decrease, increase, and slightly increase the
discrepancy in $R^{\nu}$ relative to the prediction, giving an updated
deviation of $\Delta R^{\nu} = -0.0038 \pm 0.0013$~\cite{nutev}.
Further updates are expected to incorporate QED corrections to the
PDFs and charm-mass effects.

\subsection{Yukawa Couplings (Top Mass)}
\label{sec:yukawa}

The Yukawa Higgs--fermion couplings determine the masses of all the
fermions:
\begin{equation}
\label{eq:yukawa}
{\cal{L}}_{\rm Y} = -i\sum_f \frac{g\,m_f}{2m_W}\bar{f}fH\,.
\nonumber
\end{equation}
\noindent
Of particular interest is the top quark Yukawa coupling, which is much
larger than all other Yukawa couplings and has a natural value close
to one.
The large top mass results in significant radiative corrections to the
gauge boson masses, and must be precisely known to use radiative
corrections to constrain the Higgs mass.

\begin{wrapfigure}{l}{0.5\columnwidth}
\centerline{\includegraphics[width=0.5\columnwidth]{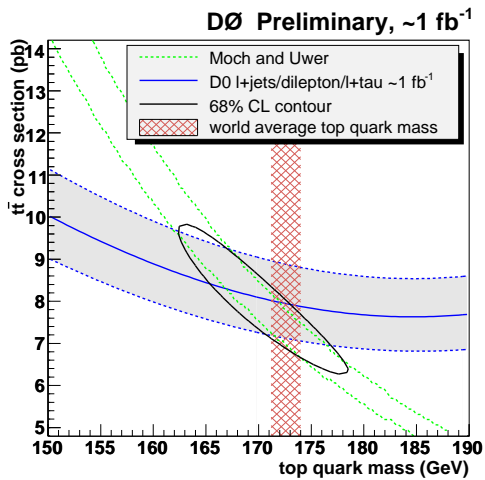}}
\vskip -0.1in
  \caption{The predicted and measured $t\bar{t}$ cross section as a
  function of top quark mass.  D\O\ uses the mass dependence of the
  cross section to extract $m_t = 169.6^{+5.4}_{-5.5}$~GeV. }
\label{fig:mtxsec}
\end{wrapfigure}

The top quark has only been observed at the Tevatron, and both CDF and
D\O\ continue to update their measurements of the top mass.
The measurements predominantly rely on reconstructing the top quarks
in $t\bar{t}$ decays to $l\nu l\nu bb$ (dilepton + jets), $l\nu qq bb$
(lepton + jets), and $qqqqbb$ (all--hadronic).
The most precise result is in the lepton + jets final state, where the
light--flavor quarks are constrained to the $W$ boson mass, reducing
the jet energy scale uncertainty.
The latest combined CDF and D\O\ result gives $m_t = 172.4 \pm
1.2$~GeV~\cite{topmass}.
The LHC experiments expect to achieve $\delta m_t = 1$~GeV with their
first fb$^{-1}$ of data.


An independent determination of the mass can be obtained through the
$t\bar{t}$ cross section measurement, since the cross section has a
significantly decreasing slope as a function of mass.
D\O\ has used its precise measurement of $\sigma_{t\bar{t}} = 7.62 \pm
0.85$~pb to extract $m_t = 170 \pm 7$~GeV \cite{d0mtxsec}.
Recently, D\O\ has compared the result to two new theoretical
calculations, one based on NLO+NLL soft--gluon
summation~\cite{ttxsecnlonll} and the other including all soft--gluon
induced NNLL terms contributing at NNLO~\cite{ttxsecnnll}.
D\O\ has also combined cross section measurements in the lepton +
jets, dilepton + jets, and lepton + tau + jets channels, obtaining its
most precise mass determination based on the cross section, $m_t =
169.6^{+5.4}_{-5.5}$~GeV, as illustrated in Figure~\ref{fig:mtxsec}.

\subsection{Kinetic Terms (Higgs Boson)}
\label{sec:kinetic}

Of the terms in the kinetic Lagrangian,
\begin{eqnarray*}
{\cal{L}}_{\rm K} & =  & \sum_f \bar{f}(i\partial \! \! \! \! \! \; / - m_f)f - \frac{1}{4}A_{\mu\nu}A^{\mu\nu}
        -\frac{1}{2}W^+_{\mu\nu}W^{-\mu\nu} + m_W^2 W^+_{\mu}W^{-\mu} - \\ 
        & & \frac{1}{4}Z_{\mu\nu}Z^{\mu\nu} + \frac{1}{2}m_Z^2 Z_{\mu}Z^{\mu} + 
                \frac{1}{2}(\partial^{\mu}H)(\partial_{\mu}H) - 
                \frac{1}{2}m_H^2H^2, \nonumber
\end{eqnarray*}
\noindent
only those associated with the Higgs boson have yet to be tested.
Direct searches from LEP have constrained the Higgs mass $m_H$ to be
greater than $114$~GeV at $95\%$ confidence level (C.L.), and an
indirect constraint from a global fit to electroweak data requires
$m_H < 160$~GeV with a best fit of $m_H = 87^{+36}_{-27}$~GeV (using
$m_t = 172.6 \pm 1.4$~GeV).
The LHC experiments expect to measure $m_W$ to a precision of $\approx
5$ MeV.
Combined with the anticipated precision in the top mass measurement of
$\approx 1$~GeV, the indirect constraint on the Higgs mass should
reach a precision of about $15$~GeV at the LHC.

\begin{wrapfigure}{!r}{0.6\columnwidth}
\centerline{\includegraphics[width=0.6\columnwidth]{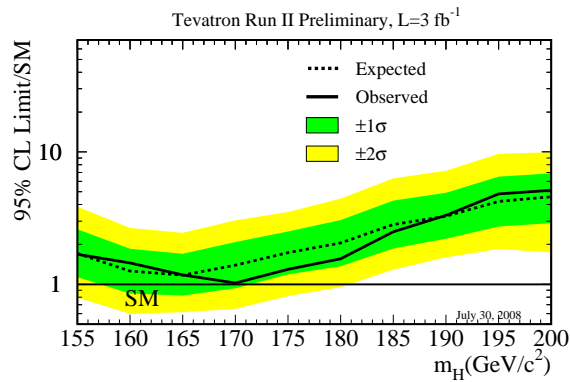}}
  \caption{The ratio of excluded to SM cross section as a function of
  Higgs boson mass. A Higgs boson with mass equal to $170$~GeV is
  excluded. }
\label{fig:tevhiggs}
\end{wrapfigure}

In the Higgs mass range allowed by the electroweak precision data, the
Tevatron has the potential to exclude or obtain evidence for the SM
Higgs with a complete data set of $\approx 7$~fb$^{-1}$.
In the low--mass region ($m_H \lesssim 130$~GeV), where the Higgs
dominantly decays to $b\bar{b}$, the searches focus on $WH$ and $ZH$
production.
A recent gain in sensitivity has come from including the $H\rightarrow
\tau\tau$ decay, with a branching ratio about $10\%$ that of $b\bar{b}$
and with lower background.
In the high--mass region ($m_H \gtrsim 130$~GeV), $H\rightarrow WW$ is
the dominant decay mode, and sensitivity comes primarily from a search
for $gg\rightarrow H \rightarrow WW \rightarrow l\nu l' \nu '$.
A recent improvement in all decay channels has come from including the
vector--boson--fusion production mechanism~\cite{newhiggs}.
The CDF and D\O\ experiments have announced the exclusion of a Higgs
with $m_h = 170$~GeV, as shown in Figure~\ref{fig:tevhiggs}.
With the data continuing to be collected, the exclusion region will
expand quickly and could turn into a claim for evidence of Higgs
production.


While the low--mass region is challenging for the LHC experiments, an
updated sensitivity study shows that discovery can be achieved with
just 5~fb$^{-1}$ for every Higgs mass by combining CMS and ATLAS
searches.

\section{Beyond the Standard Model Searches}
\label{sec:searches}

\subsection{Searches for Rare Processes at High Transverse Momentum}

\subsubsection{General Searches for New Physics}

\begin{wrapfigure}{r}{0.5\columnwidth}
  \centerline{\includegraphics[width=0.65\columnwidth, angle=270]{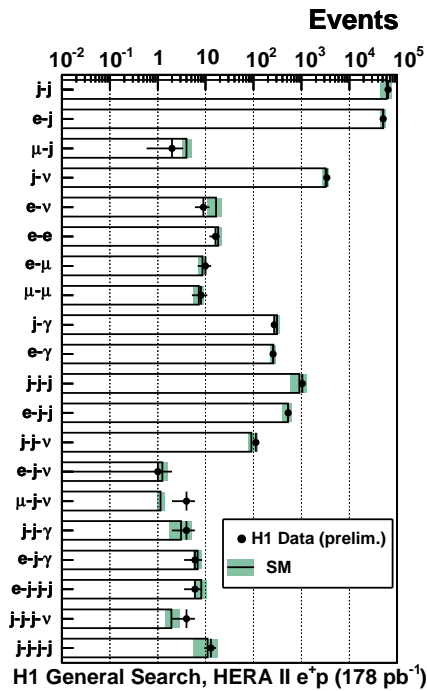}}
  \caption{The data and SM expectation for all event classes with observed
    data events or a SM expecation greater than 1 event in the H1
    general search analysis of the HERA~II $e^{+}p$ data.
    The error bands on the predictions include model uncertainties and
    experimental systematic errors added in quadrature.
}
\label{fig:generalSearch}
\end{wrapfigure}

The H1 Collaboration has performed a generic search for deviations
from the SM at large values of transverse momentum $P_{T}>20$~GeV in
the full H1 data set of  $0.5$~fb$^{-1}$~\cite{h1general}.
All high $P_{T}$ final--state configurations involving electrons ($e$),
muons ($\mu$), jets ($j$), photons ($\gamma$) or neutrinos ($\nu$) are
systematically investigated in exclusive classes, formed according to
the number and types of objects detected in the final state
(e.g. $e$-$j$, $\mu$-$j$-$\nu$, $j$-$j$-$j$-$j$-$j$). 
Event yields are shown in Figure~\ref{fig:generalSearch} for the H1
HERA~II $e^{+}p$ data, where no significant deviation from the SM is
observed in the phase space and event topologies covered by the
analysis.

A similar global search for new physics based on a systematic search
for discrepancies with respect to the SM has also been performed by
CDF using $2$~fb$^{-1}$ of high $P_{T}$ data, where $399$ exclusive final
states and almost $20000$ kinematic distributions are searched~\cite{cdfgeneral}.
Although a number of interesting effects are observed, the search
reveals no indication of physics beyond the Standard Model.

\subsubsection{Events with High--$P_{T}$ Leptons}

\begin{figure}[tp]
  \includegraphics[width=.47\textwidth]{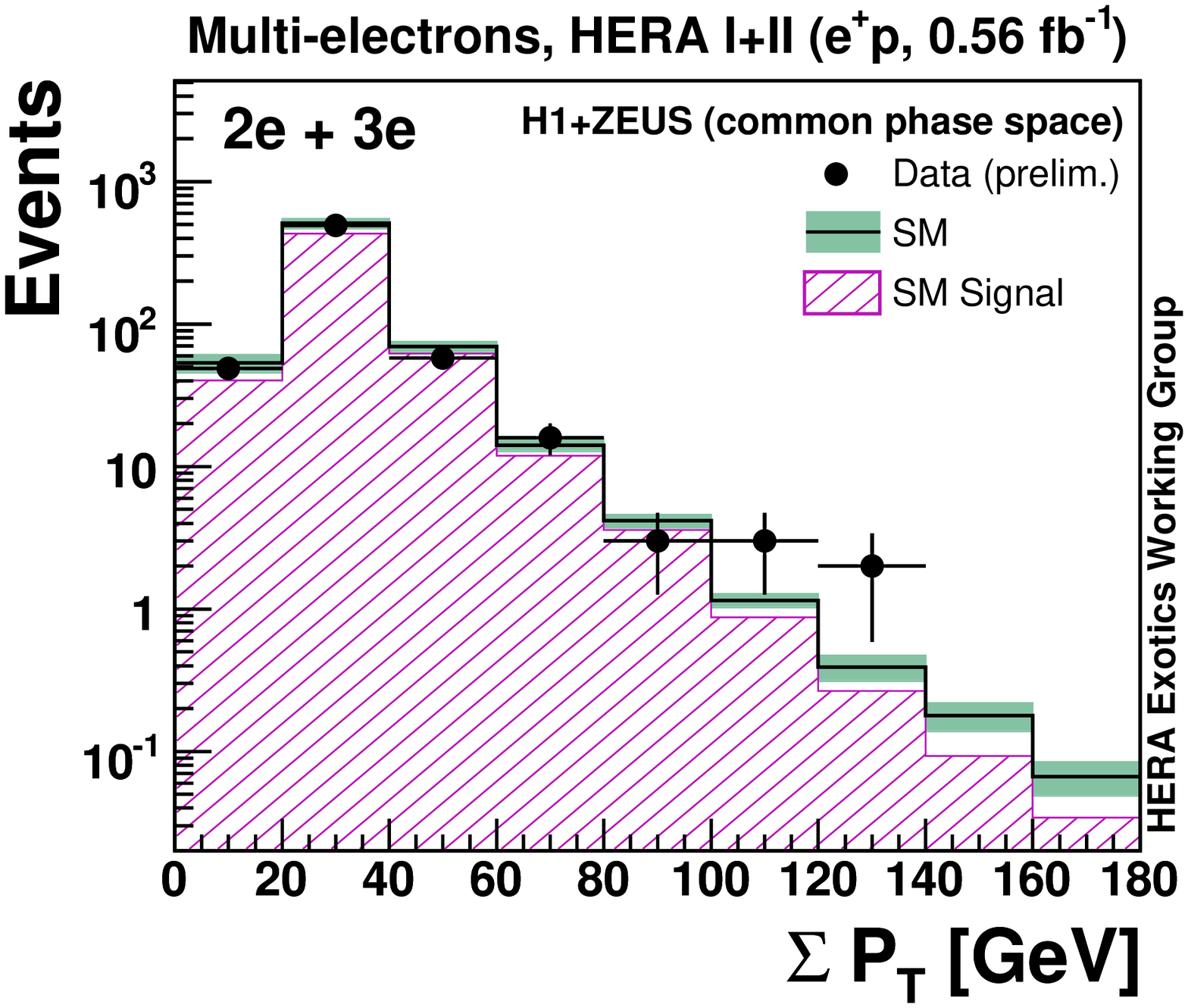}
  \hspace{1.2cm}
  \includegraphics[width=.42\textwidth]{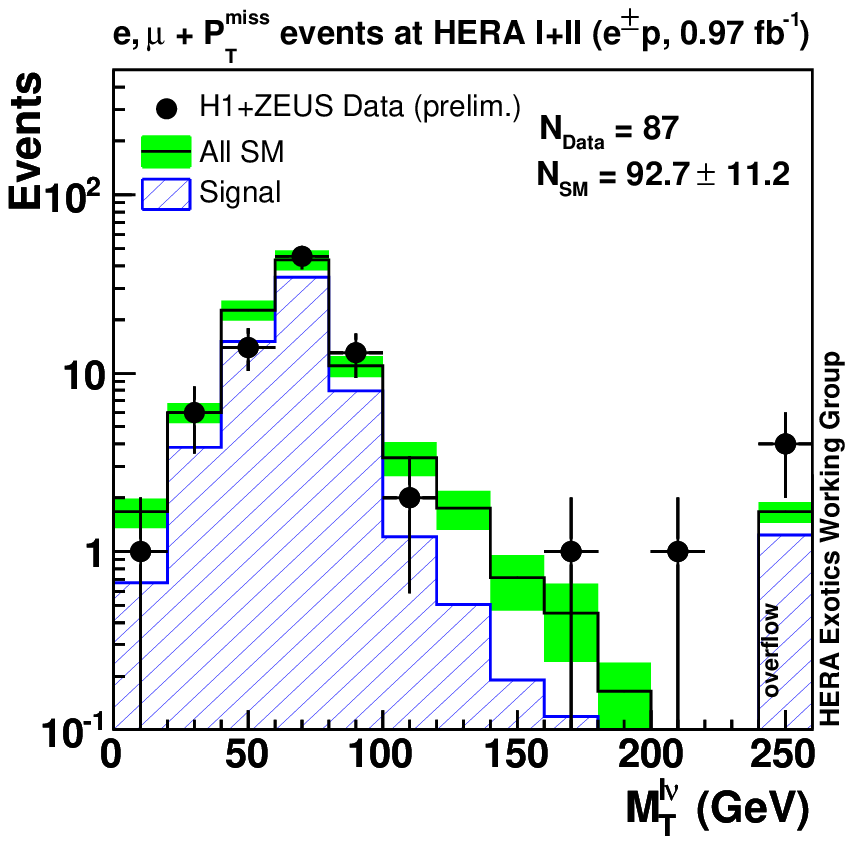}
    \caption{Left: The scalar sum of the electron transverse momenta
    for events with $2$ or $3$ electrons in the combined H1+ZEUS
    $e^{+}p$ data. Right: The transverse mass $M_{T}^{l\nu}$
    distribution for events with the isolated lepton and missing
    transverse momentum in the combined H1+ZEUS $e^{\pm}p$ data.  The
    data (points) are compared to the SM expectation (open
    histogram). The signal component of the SM expectation (left:
    photon--photon interactions, right: $W$ production) is given by
    the hatched histogram. $\rm N_{Data}$ is the total number of data
    events observed and $\rm N_{SM}$ is the total SM expectation. The
    total uncertainty on the SM expectation is given by the shaded
    band.}
\label{fig:highPtLeptons}
\end{figure}

Measurements of final states with multiple high $P_{T}$ electrons and
muons are performed by H1 and ZEUS~\cite{multilep}.
Within the SM the production of multi--lepton events
in $ep$ collisions mainly proceeds via photon--photon
interactions.
The observed event yields are in good agreement with SM expectations,
although in the H1 analysis an excess of data events is observed at
high masses in the $e^{+}p$ data.
A common phase space has been established in order to combine the
results of the two experiments, using a total integrated
luminosity of $0.94$~fb$^{-1}$~\cite{multilepcombined}.
As an example, for multi--electron events the distribution of the
scalar sum of all electron transverse momenta $\sum P_{T}$ is shown in
Figure~\ref{fig:highPtLeptons}~(left) for 2 and 3 electron events in
the $e^{+}p$ data.
For $\sum P_{T} >100$~GeV a slight excess is present with 5 events
observed in the data, compared to a SM expectation of $1.8\pm0.2$.
No such excess is seen in the $e^{-}p$ data.

Searches for events containing isolated leptons (electrons, muons or
tau leptons) in coincidence with large missing transverse momentum are
also performed by the H1 and ZEUS Collaborations~\cite{isolep}.
Such events occur at HERA within the SM via $W$ production with
subsequent leptonic decay, which has a cross section of order $1$~pb.
%
%
The full HERA data has been analysed by both H1 and ZEUS.
The latest electron and muon results from H1 reveal no excess over the
SM for the $e^{-}p$ data. However, the previously observed excess over
SM prediction persists in the $e^{+}p$ data at large hadronic
transverse momentum, $P_{T}^{X}>25$~GeV, where $21$ data events are
observed compared to an expectation of $8.9\pm1.5$.
The electron and muon results from ZEUS show no excess over the SM
prediction.
Similarly, H1 observes good agreement with the SM in its search for
events with isolated tau leptons and missing $P_{T}$.
H1 and ZEUS have combined their data in the electron and muon
channels, resulting in a data set with a total integrated luminosity
of $0.97$~fb$^{-1}$~\cite{isolepcombined}.
The distribution of the reconstructed transverse mass $M_T^{\ell\nu}$
of the H1 and ZEUS events in the complete $e^{\pm}p$ HERA data is
shown in Figure~\ref{fig:highPtLeptons}~(right), where a clear
Jacobian peak is observed, consistent with the expectation from SM $W$
production.


Events containing isolated leptons and large missing transverse
momentum would be produced at HERA in several models of physics beyond
the SM such as single top production via flavour--changing neutral
current (FCNC)~\cite{h1top}.
In particular, this process would produce events with large values of
$P_{T}^{X}$, whereas SM $W$ production is expected to predominantly
produce events with small $P_{T}^{X}$.
The H1 events are interpreted in this model, and using a maximum
likelihood method an upper limit on the anomalous top production cross
section of $\sigma_{ep\rightarrow etX}~<~0.16$~pb is established at
$95\%$~C.L., corresponding to a limit on the anomalous coupling
$\kappa_{tu\gamma}<0.14$, which is currently the world's most
stringent limit.
The CDF Collaboration performs a search for the FCNC decay of the top
quark $t \rightarrow Zq$ using $1.9$~fb$^{-1}$ of Tevatron Run~II
data~\cite{cdffcnc}.
%
%
$Z$ + $\geq 4$ jets candidate events are used, both with and without a
secondary vertex $b$--tag, and the background is rejected using kinematic
constraints present in FCNC events.
In the absence of a signal an upper limit on the branching fraction of
$B(t \rightarrow Zq)<3.7\%$ at $95\%$~C.L. is obtained, which is
currently the world's best limit on the anomalous vector coupling $V_{tuZ}$.

\subsection{Searches for New Particles and Phenomena}

\subsubsection{Searches for High--Mass Particles}

\begin{figure}[htb]
  \includegraphics[width=.43\textwidth]{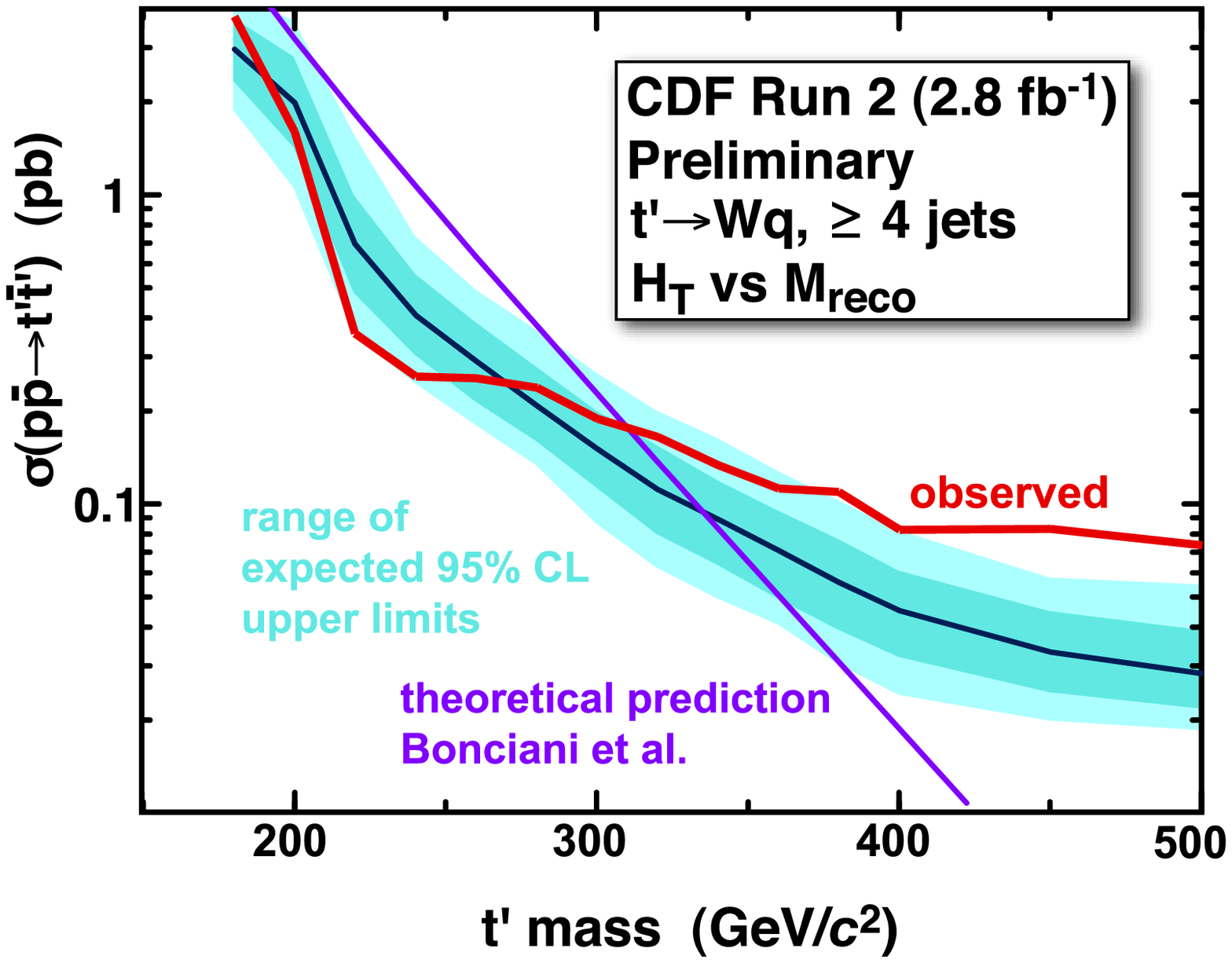}
  \hspace{0.5cm}
  \includegraphics[width=.50\textwidth]{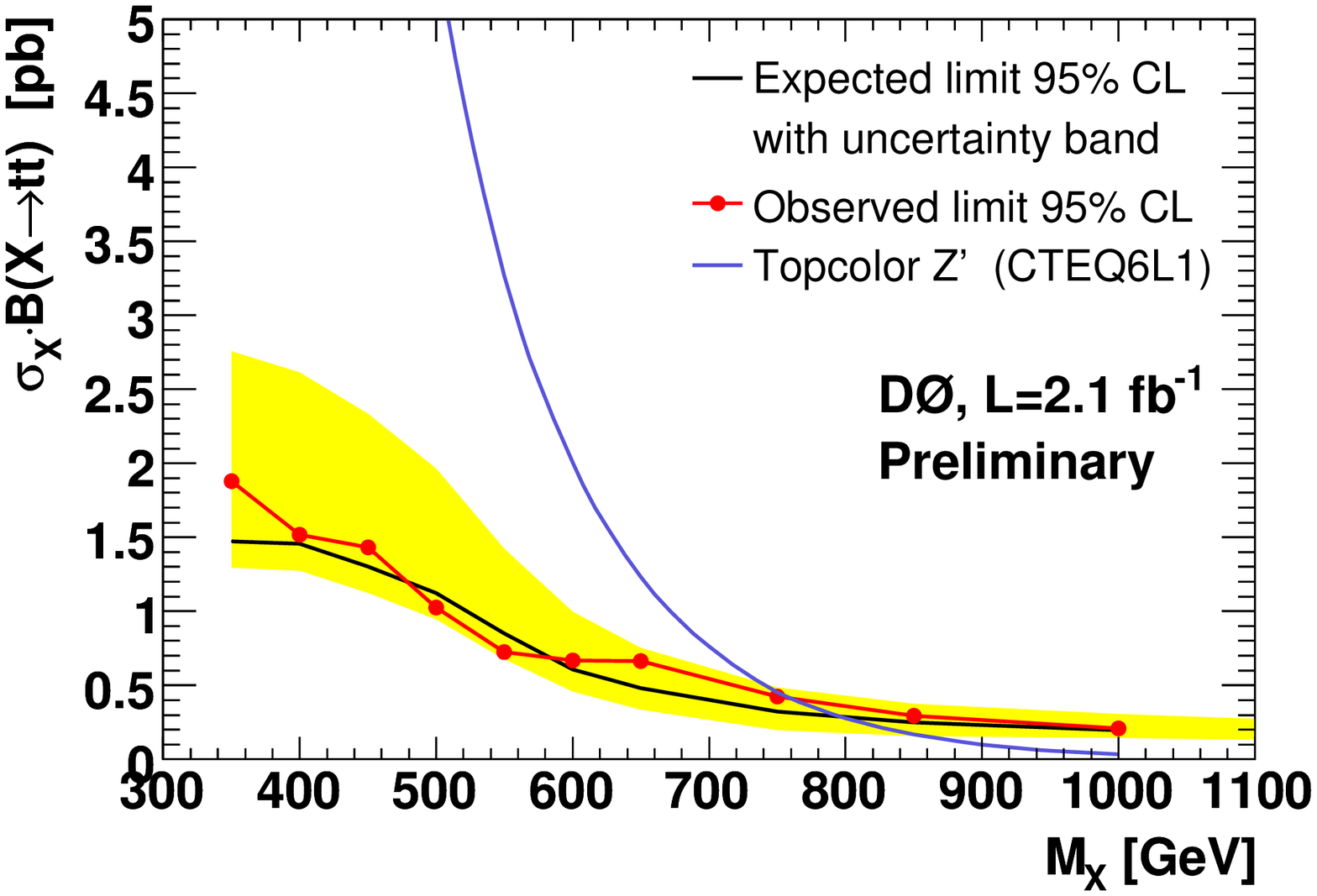}

    \caption{Left: Upper limit at $95\%$~C.L. from CDF on the cross
    section for $t'$ pair production at the Tevatron as a function of
    $t'$ mass (red). The purple curve is the theoretical cross
    section. The blue band represents the range of limits expected
    from $\pm 1\sigma$ fluctuations in the predicted background (light
    blue corresponds to $\pm 2\sigma$). Right: Expected and observed
    $95\%$~C.L. upper limits from D{\O} on $\sigma_X\cdot
    B(X\rightarrow t\bar{t})$ compared with the predicted
    top--colour--assisted technicolour cross--section for a $Z'$ boson
    with a width of $\Gamma_{Z'}=0.012~M_{Z'}$ as a function of the
    resonance mass $M_X$. The shaded band gives the range of expected
    limits corresponding to $\pm 1\sigma$ background fluctuations.}

\label{fig:heavyBoson}
\end{figure}

The CDF Collaboration performs a search for a heavy top--like object
$t'$ using $2.8$~fb$^{-1}$ of Tevatron Run~II data~\cite{cdftprime}.
A large $t'$ branching ratio to $Wq$ is assumed, as would be the case
if $M_{t'} < M_{b'} + M_W$, a situation favoured by the constraint
that an additional quark generation be consistent with precision
electroweak data.
The search is performed using two kinematic variables to separate the
$t'$ signal from the SM background: $H_T$, the sum of the transverse
momenta of all objects in the event, and $M_{\rm reco}$, the $Wq$
reconstructed invariant mass.
No evidence for $t'$ is observed and an upper limit on
$\sigma(t'\bar{t'})$ at $95\%$~C.L. is set, excluding $t'$ masses
below $311$~GeV, as shown in Figure~\ref{fig:heavyBoson}~(left).

The D{\O} Collaboration performs a generic search for a narrow resonance
decaying to top pairs $X\rightarrow t\bar{t}$, using $2.1$~fb$^{-1}$
of Tevatron Run~II data~\cite{d0ttprime}.
The width $\Gamma$ is assumed to be small compared to the detector
mass resolution.
The signal topology investigated is $3$ or $4$ jets plus a lepton, and
missing transverse energy.
%
%
No significant deviation from the SM expectation is observed in the
reconstructed $M_{t\bar{t}}$ distribution.
An upper limit on the production cross section $\sigma_X\cdot
B(X\rightarrow t\bar{t})$ is derived as a function of resonance mass
$M_X$, as shown in Figure~\ref{fig:heavyBoson}~(right).
A model for $Z'$ production is also shown, for which corresponding $Z'$
masses $M_{Z'}<760$~GeV are ruled out at $95\%$~C.L.


Also performed by D{\O} and CDF are
searches for heavy resonances decaying to fermion pairs in the di--jet,
di--electron, four--electron, and electron plus missing transverse
energy channels.
No significant excess is observed in the data in all cases, and
improved limits are set on the production of such particles, often
extending well beyond $1$~TeV, as is the case of the colour--octet
techni--$\rho$ (up to $1.1$~TeV) and axigluon and flavour--universal
colouron (up to $1.25$~TeV) models~\cite{cdfpairs}.
%

\subsubsection{Excited Fermion Searches}

The existence of excited states of leptons and quarks is a natural
consequence of models assuming composite fermions, and their discovery
would provide convincing evidence of a new scale of matter.
The production and decay of such particles is described in
gauge--mediated (GM) and contact--interaction (CI) models.

\begin{figure}[tbh]
  \includegraphics[width=.46\textwidth]{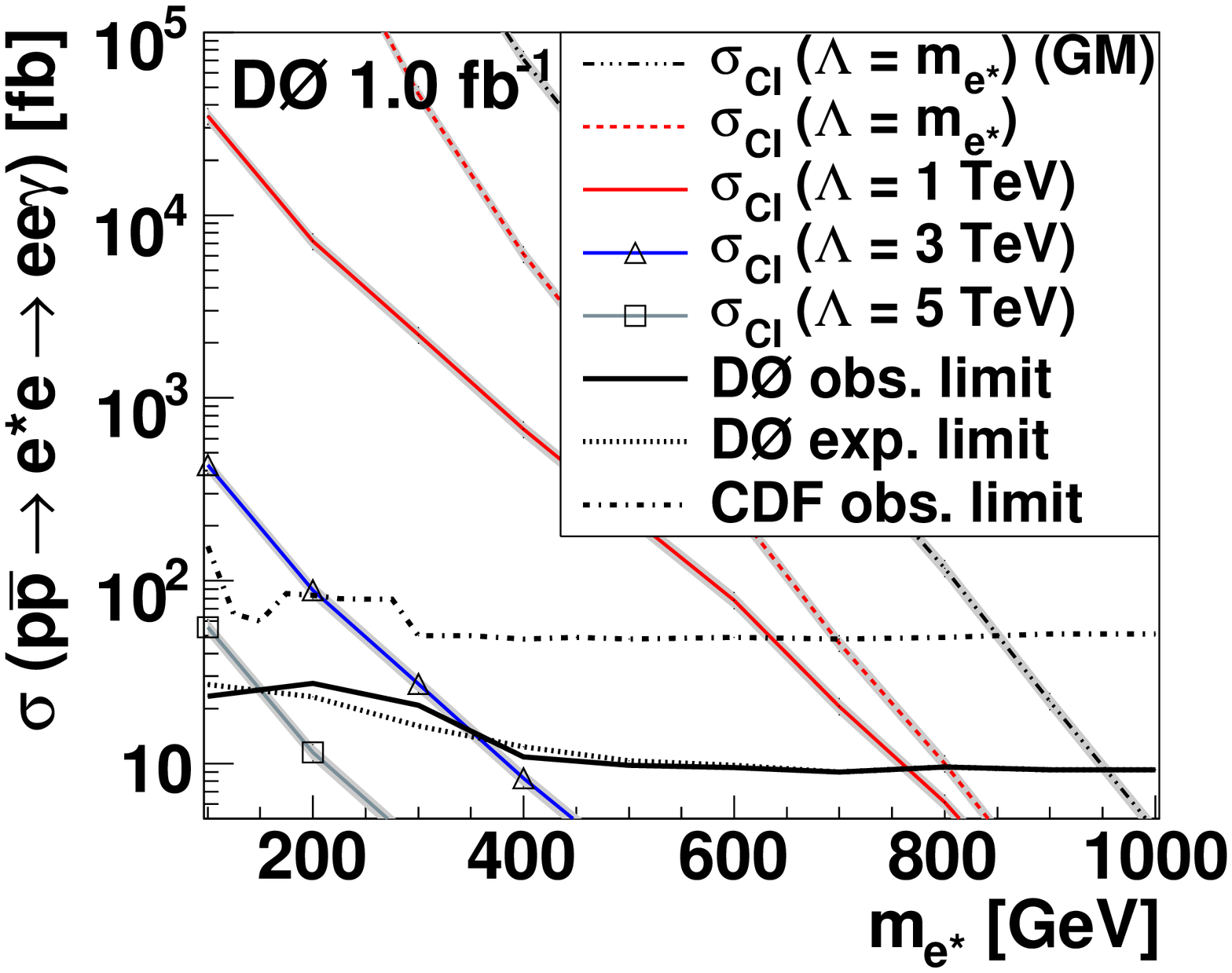}
  \hspace{0.7cm}
  \includegraphics[width=.44\textwidth]{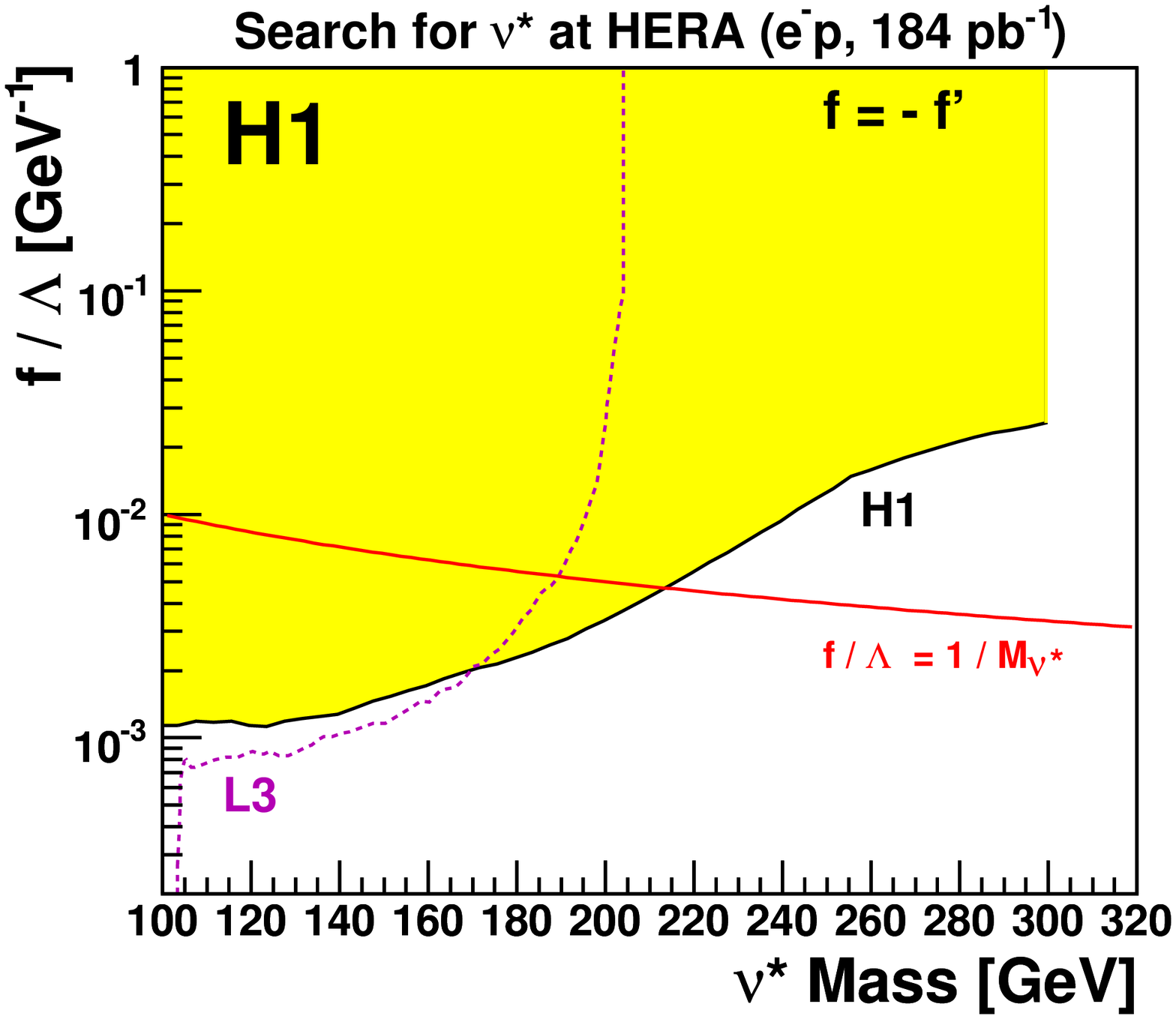}
    \caption{Left: The D{\O} measured and expected limits on cross
    section times branching fraction, compared to the CI model
    prediction for different choices of $\Lambda$. Also shown is the
    prediction under the assumption that no decays via contact
    interactions occur (GM). The observed limit from CDF is also
    indicated.  Right: H1 exclusion limit at $95\%$~C.L. on the
    coupling $f/\Lambda$ as a function of the mass of the excited
    neutrino with the assumption $f = -f'$. The excluded domain, based
    on all H1 $e^{-}p$ data, is represented by the shaded area. The
    dashed line corresponds to the exclusion limit obtained at LEP by
    L3.}
\label{fig:excitedFermions}
\end{figure}

D{\O} has searched for $e^{*}$ in the process $p \bar{p} \rightarrow
e^{*}e$, with subsequent $e^{*}$ decay to an electron plus a
photon~\cite{d0excited}.
Events with two isolated high $P_{T}$ electrons and one isolated high
$P_{T}$ photon are selected in $1$~fb$^{-1}$ of data, where the SM
background is dominated by the Drell--Yan process, DY + $\gamma
\rightarrow e^+e^- \gamma $.
No excess is seen in the data, and therefore $95\%$~C.L. limits are
derived, including both CI and GM decays.
The resulting limits are shown as a function of $m_{e^*}$ in
Figure~\ref{fig:excitedFermions}~(left), together with predictions of
the CI model for different choices of the compositeness scale
$\Lambda$.
For $\Lambda = 1$~TeV, masses below $756$~GeV are excluded.
The excited electron decay channels ${e}^{*} {\rightarrow}
{e}{\gamma}$,  ${e}^{*} {\rightarrow} {e}{Z}$ and ${e}^{*}
{\rightarrow} {\nu}{W}$ with subsequent hadronic or leptonic decays of
the $W$ and $Z$ bosons are examined at HERA in a search for excited electrons
by H1~\cite{h1excitedelec}.
In this search, which uses the complete H1 $e^{\pm}p$ data sample 
 of $475$~pb$^{-1}$ integrated luminosity, no indication of a signal is
observed.
New limits on the production cross section of excited electrons are
obtained within a GM model, where an upper limit on the
coupling $f/\Lambda$ as a function of the excited electron mass is
established for the specific relation $f = +f'$ between the couplings.
Assuming $f = + f'$ and $f/\Lambda=1/M_{e^*}$ excited electrons with a
mass lower than $272$~GeV are excluded at $95\%$~C.L.
For the first time in $ep$ collisions, gauge and four--fermion contact
interactions are also considered together for $e^*$ production and
decays, although it is found that the CI term improves the limit on
$1/\Lambda$ only slightly, demonstrating that the GM mechanism is
dominant for excited electron processes at HERA.

Using the full $e^{-}p$ data sample collected by the H1 experiment at
HERA with an integrated luminosity of $184$~pb$^{-1}$ a search for the
production of excited neutrinos is performed~\cite{h1excitednu}.
Due to the helicity dependence of the weak interaction and given the
valence quark composition and density distribution of the proton, the
$\nu^{*}$ production cross section is predicted to be much larger
for $e^-p$ collisions than for $e^+p$.
The excited neutrino decay channels ${\nu}^{*} {\rightarrow}
{\nu}{\gamma}$,  ${\nu}^{*} {\rightarrow} {\nu}{Z}$ and ${\nu}^{*}
{\rightarrow} {e}{W}$ with subsequent hadronic or leptonic decays of
the $W$ and $Z$ bosons are considered and no indication of a $\nu^*$
signal is found.
Upper limits on the coupling $f/\Lambda$ as a function of the excited
neutrino mass are established for specific relations between the
couplings.
Assuming $f = - f'$ and $f/\Lambda=1/M_{\nu^*}$, excited neutrinos
with a mass lower than $213$~GeV are excluded at $95$\%~C.L., as
shown in Figure~\ref{fig:excitedFermions}~(right).
The excluded region is greatly extended with respect to previous results
and demonstrates the unique sensitivity of HERA to excited neutrinos
with masses beyond the LEP reach.

\subsubsection{Searches for Leptoquarks}

\begin{wrapfigure}{l}{0.45\columnwidth}
  \centerline{\includegraphics[width=0.5\columnwidth]{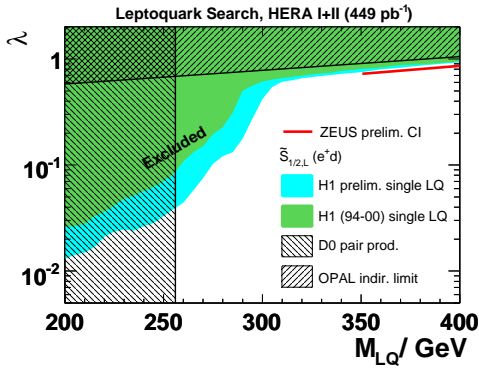}}
\vskip -0.1in
  \caption{H1 exclusion limits at $95\%$~C.L. on the coupling
    $\lambda$ as a function of the LQ mass for the $\tilde{S}_{1/2,L}$
    leptoquark in the framework of the BRW model. Corresponding
    limits from LEP (OPAL) and the Tevatron (D{\O}) are also shown, as
    well as the indirect CI limits from ZEUS (see section \ref{Sec:CI}).} 
\label{fig:leptoquark}
\end{wrapfigure}
A search for leptoquark (LQ) production is performed by H1 at HERA in
the 14 LQ framework of the Buchm\"uller, R\"uckl and Wyler (BRW)
model, using the polarised $e^{\pm}$ HERA~II data~\cite{h1lq}.
LQs may be resonantly produced at HERA up to the centre--of--mass
energy, beyond which the production mechanism is CI--like.
An irreducible SM background is present from neutral and charged
current DIS.
No signal is observed in the DIS mass spectra and constraints on
LQs are set, which extend beyond the domains previously excluded.
For a coupling of electromagnetic strength, LQ masses below
$291$--$330$~GeV are ruled out at HERA, depending on the LQ type.
In a separate search of the second--generation LQs at HERA,
no signal for the lepton flavour violating (LFV) process
$ep \rightarrow \mu X$ is found in the complete H1 $e^{-}$
HERA~II data set~\cite{h1lfv}.
For Yukawa couplings of electromagnetic strength, LQs mediating the
LFV process $e \rightarrow \mu$ are ruled out for LQ masses
up to $433$~GeV.
%


In $p\bar{p}$ collisions, LQs would be produced in pairs via the
strong interaction.
The production rate is thus essentially independent of the unknown Yukawa 
couplings.
%
%
The most recent LQ searches at the Tevatron have concentrated on
$jj\nu\nu$ and $jj\mu\nu$ final states, more details of which can be
found in~\cite{beauchemin}.
Table~\ref{Tab:tevatronLeptoquarks} summarises all of the latest
limits obtained by CDF and D{\O} for the different final states to
which the different types of LQ could contribute.
This table represents all the constraints for each of the three
generations of LQs, imposed by the Tevatron on a generic LQ model.

\begin{table}[t]
\renewcommand{\arraystretch}{1.2}
\centerline{\begin{tabular}{|c||c|c|c||c|c|c|}
\hline
& \multicolumn{3}{c||}{CDF} & \multicolumn{3}{c|}{D{\O}}\\
\hline
\hline
Final state    & 1$^{\mbox{st}}$ gen. & 2$^{\mbox{nd}}$ gen. & 3$^{\mbox{rd}}$ gen.  & 1$^{\mbox{st}}$ gen. & 2$^{\mbox{nd}}$ 
gen. & 3$^{\mbox{rd}}$ gen.   \\\hline
$\ell\ell jj$  &     236 GeV    &     225 GeV    &     151 GeV    &     256 GeV    &     251 GeV    &     180 GeV    \\\hline
$\ell\nu jj$   &     205 GeV    &     208 GeV    &        -       &     234 GeV    &     214 GeV    &      -         \\\hline
$\nu\nu jj$    &     177 GeV    &     177 GeV    &     167 GeV    &     136 GeV    &     136 GeV    &     229 GeV    \\\hline
\end{tabular}}
  \caption{Summary of the $95\%$~C.L. limits set by CDF and D{\O} on
  LQ masses of each generation, from searches using different final
  states containing 2, 1 or 0 charged leptons $\ell$.}
\label{Tab:tevatronLeptoquarks}
\end{table}

\subsubsection{Contact Interactions}
\label{Sec:CI}

\begin{wrapfigure}{r}{0.45\columnwidth}
  \vspace*{-10mm}
  \centerline{\includegraphics[width=0.43\columnwidth,scale=0.85]{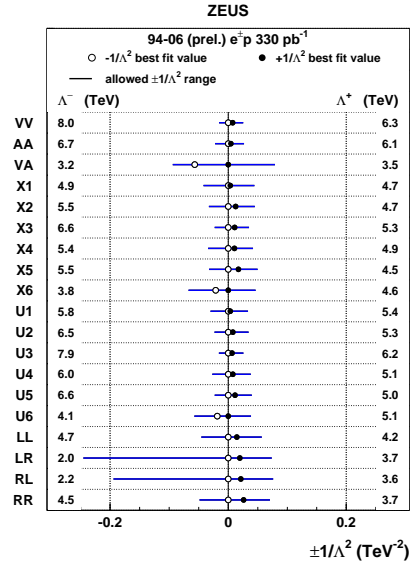}}
    \caption{Results for general contact interaction models
    (compositeness models) obtained using the combined $e^{+}p$ and
    $e^{-}p$ data from ZEUS (1994-2006). Horizontal bars indicate the
    $95\%$ C.L. limits on $\eta / 4\pi = \varepsilon / \Lambda^{2}$
    and values outside these regions are excluded. $\Lambda^{\pm}$ are
    the $95\%$ C.L. limits on the compositeness scale for $\varepsilon
    = \pm 1$.}
\label{fig:CI}
\end{wrapfigure}
New interactions between electrons and quarks at HERA involving mass
scales above the centre--of--mass energy can modify the deep inelastic
$e^{\pm}p$ scattering cross sections at high $Q^{2}$ via virtual effects,
resulting in observable deviations from the SM predictions.
Four--fermion contact interactions are an effective theory, which
allows a general description of their effects.
Various scenarios are considered by H1~\cite{h1CI} and
ZEUS~\cite{zeusCI}.
In the general case (also referred to as compositeness models), 
limits on the effective ``new physics'' mass scale $\Lambda$ 
(the compositeness scale) are extracted assuming that the 
coefficients $\eta$ of the various terms in the effective 
Lagrangian are given by  
$\eta = \pm 4\pi / \Lambda^{2}$.
Figure~\ref{fig:CI} shows the results obtained by ZEUS for different
compositeness models, based on the analysis of 1994--2006 data.
Limits on the effective mass scale $\Lambda$ range from $2$--$8$~TeV.
%
%

For a model with large extra dimensions (LED), where cross section
deviations are expected due to a graviton--exchange contribution, ZEUS
set limits on the effective Planck mass scale $M_{S}$, and scales up
to $0.90$~TeV are excluded at $95\%$ C.L.
Direct searches for LED performed at the Tevatron in the form of
graviton production also imply $\Lambda>1.5$~TeV, with
$M_{G}>1$~TeV~\cite{krutelyov}.
Searches for possible quark substructure can be performed at HERA by
measuring the spatial distribution of the quark charge.
By using the {\it classical} form factor approximation, and assuming
that both electron and exchanged bosons are point--like, limits on the
mean--square radius of the electroweak charge of the quark are set,
where quark radii bigger than $0.74\cdot 10^{-16}$~cm (H1) and
$0.62\cdot 10^{-16}$~cm (ZEUS) are excluded at $95\%$ C.L.
Contact interactions can also be used to describe the effects of
virtual LQ production or exchange at HERA, in the limit of large LQ
mass $M_{LQ} \gg \sqrt{s}$.
Indirect LQ limits from ZEUS~\cite{zeusCI} are shown in the CI
kinematic domain in Figure~\ref{fig:leptoquark}, compared to the
direct H1 exclusion limits discussed above.

\subsection{Supersymmetry}

In the search for new phenomena, a well--motivated extension to the SM
is supersymmetry (SUSY), which relates particles with different spin.
SUSY is one of the most promising ways to solve crucial problems of the SM.
This spacetime symmetry links bosons to fermions and introduces
supersymmetric partners (sparticles) to all SM particles.

\subsubsection{SUSY Searches in the mSUGRA Model}

\begin{figure}[htb]
  \includegraphics[width=.45\textwidth]{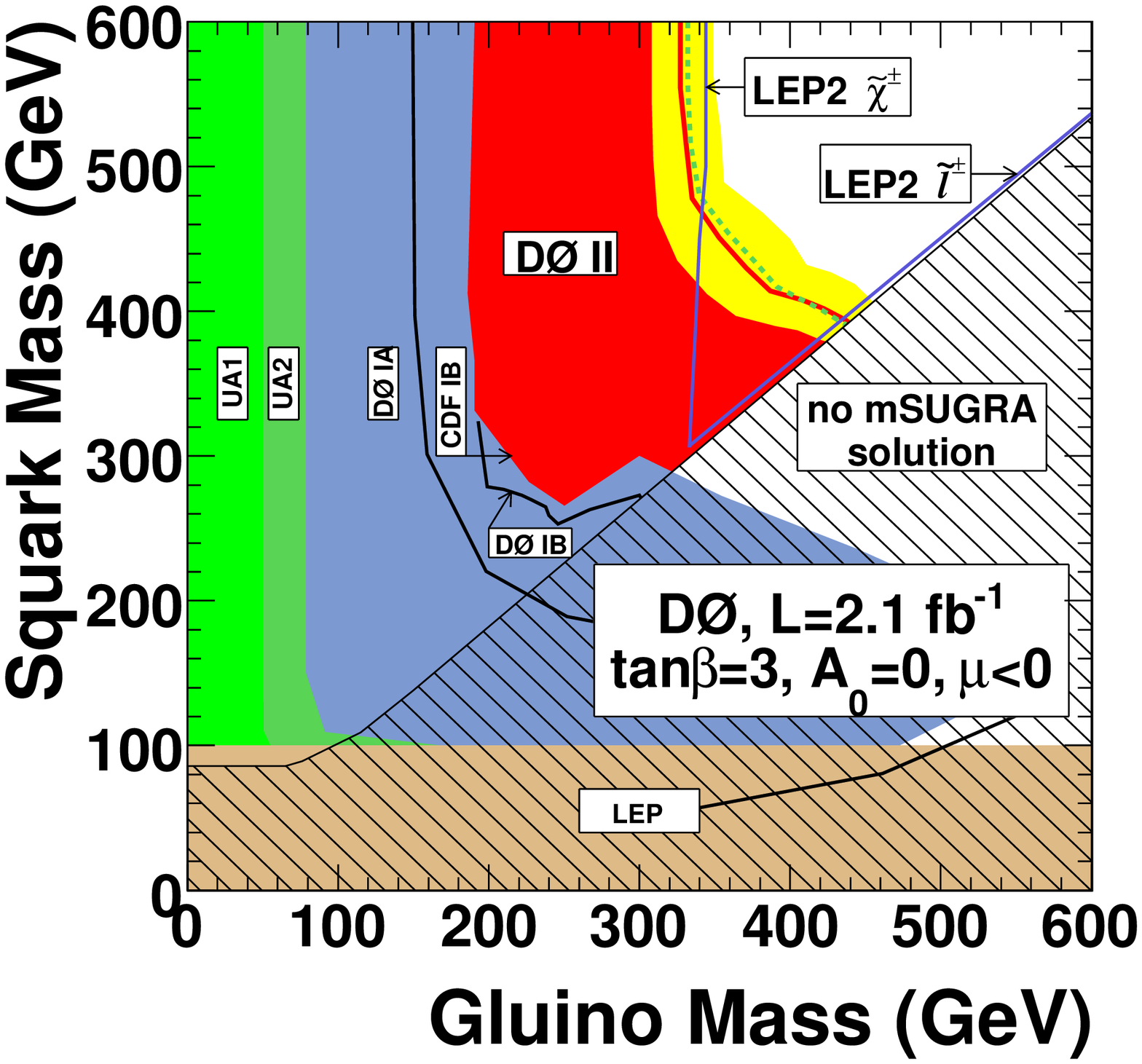}
  \hspace{0.5cm}
  \includegraphics[width=.5\textwidth]{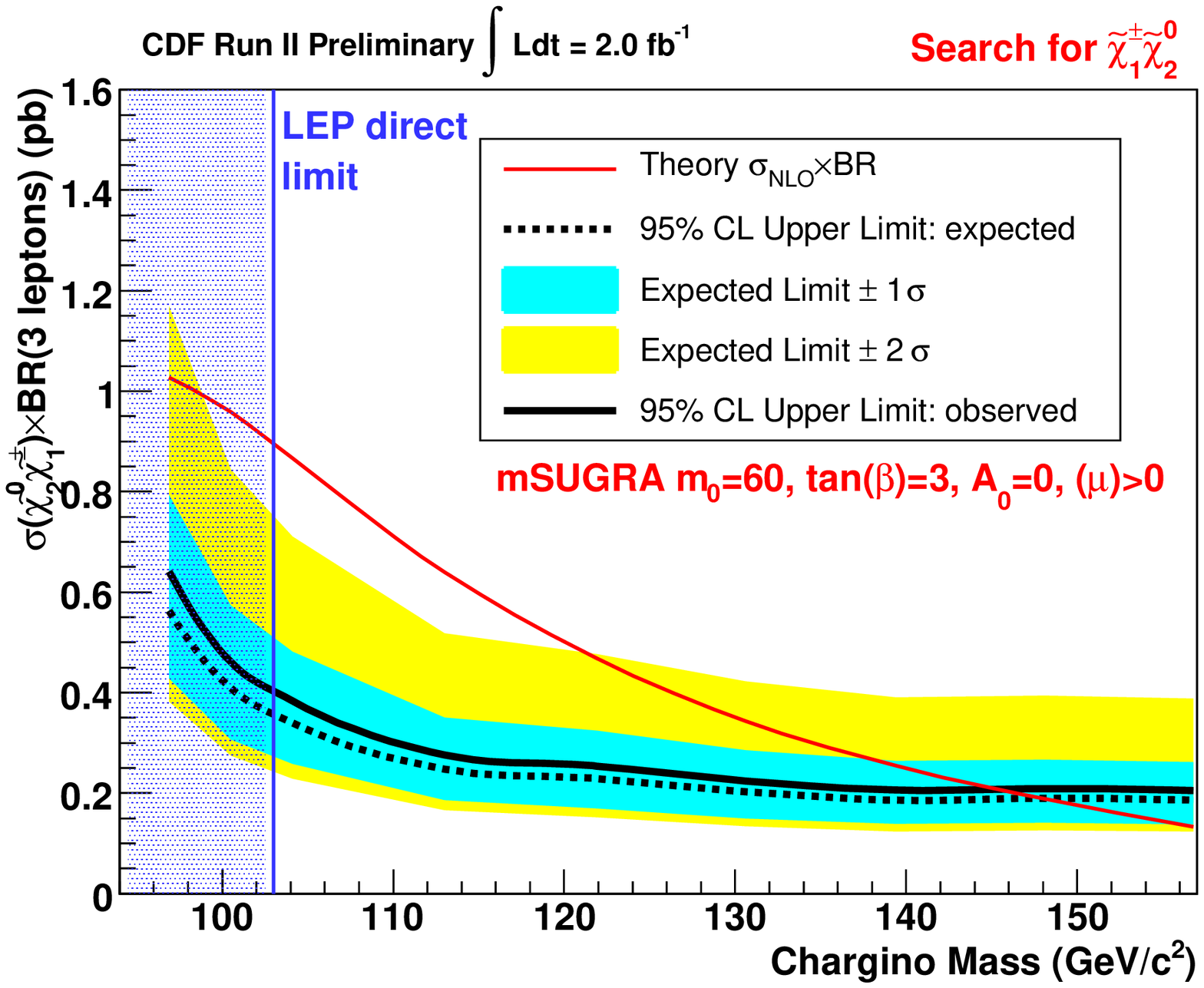}
  \caption{Left: Limits on squark and gluino masses from D{\O}. Right:
    CDF cross section times branching ratio upper limit as a
    function of chargino mass.}
\label{fig:susyLimits}
\end{figure}

Searches for squarks and gluinos are performed by the CDF and D{\O}
Collaborations, using Tevatron Run~II data~\cite{squarksandgluons}.
The searches are performed in the minimal super gravity (mSUGRA) model
with the lightest neutralino $ \tilde{\chi}_{1}^{0} $ as the lightest
supersymmetric particle (LSP).
Three different regimes are defined, leading to the analysis of three
separate final states: events with $2$ jets, $3$ jets and at least $4$
jets, all in combination with missing transverse energy.
All squark species are considered except the stop (CDF and D{\O})
and the sbottom (CDF).
The main SM background is due to $W$/$Z$ + jets, di--bosons, and
$t\bar{t}$ events.
A good agreement is observed between data and the SM expectation
in samples of $2$~fb$^{-1}$.
Lower limits are derived on the masses of squarks and gluinos, taking
into account the systematic and statistical uncertainties.
The limits from the D{\O} experiment are shown in
Figure~\ref{fig:susyLimits}~(left), where for $\tan\beta=3$,
$A_0=0$, and $\mu<0$, squark masses lower than $392$~GeV and gluino
masses lower than $327$~GeV are excluded at $95\%$ C.L.
The CDF and D{\O} results are the best constraints to date on the
squark and gluino masses.

One of the promising modes for SUSY detection at hadron colliders is
that of chargino--neutralino associated production with decay into a
trilepton signature.
Charginos decay into a single lepton through a slepton or gauge boson,
and neutralinos similarly decay to two detectable leptons. Thus the
detector signature is three SM leptons with associated missing energy
from the undetected neutrinos and lightest neutralinos,
$\tilde{\chi}_1^0$, in the event.
Due to its electroweak production, this is one of the few {\it
jet--free} SUSY signatures.
CDF performs a search using $2$~fb$^{-1}$ of data for a series of
trilepton an dilepton~$+$~track exclusive channels~\cite{charginos}.
Good agreement is observed between the data and the SM.
As no clear signal is seen, upper limits on cross section times
branching ratio are derived as a function of chargino mass, as shown
in Figure~\ref{fig:susyLimits}~(right).
A chargino lower mass limit of $145$ GeV is obtained by CDF and in a
similar analysis a lower mass limit of $145$ GeV is also obtained by
D{\O}.


As the turn--on of the LHC approaches, the discovery reach of ATLAS and
CMS at the LHC in different search channels has been investigated
using mSUGRA parameter scans~\cite{bruneliere}.
Gluino and squark masses less than $\mathcal{O}$($1$)~TeV are reported to
be within reach after having accumulated and understood an integrated
luminosity of about $1$~fb$^{-1}$.

\subsubsection{SUSY Searches in the GMSB Model}

\begin{wrapfigure}{r}{0.45\columnwidth}
  \centerline{\includegraphics[width=0.45\columnwidth]{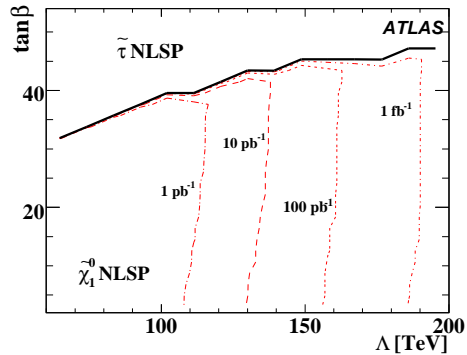}}
  \caption{Contour lines with 5 di--photon + missing transverse energy
    signal events in the $\Lambda$--$\tan\beta$ plane for different
    integrated luminosities collected by the ATLAS experiment.}
\vspace*{5mm}
\label{fig:susyLHC}
\end{wrapfigure}

In SUSY models with gauge mediated supersymmetry breaking (GMSB) the
gravitino is the lightest SUSY particle (LSP), while the lightest
neutralino or slepton is the next--to--lightest sparticle (NLSP).
GMSB models offer a possibility to break SUSY at a low scale compared to
that of generic mSUGRA models.
At the Tevatron, cascade decays of pair--produced neutralinos to LSPs
produce a final state featuring large missing transverse energy (from
the gravitinos) and two photons (or leptons). 
The D{\O} experiment has searched for such a topology using
$1$~fb$^{-1}$ of data and, in the absence of a signal, set the current
best limits on the masses of the neutralino
$M_{\tilde{\chi}_1^0}>125$~GeV, the chargino
$M_{\tilde{\chi}_1^+}>229$~GeV and the symmetry breaking scale
$\Lambda>91.5$~TeV~\cite{d0gmsb}.

In order to estimate the LHC discovery potential of the di--photon
channel, a scan of the breaking scale $\Lambda$ and $\tan\beta$ has
been performed using a fast simulation.
%
%
In this region the neutralino is usually the NLSP, except for large
$\tan\beta$.
Figure~\ref{fig:susyLHC} shows the contour lines with 5 signal events
for different integrated luminosities collected by the ATLAS
experiment after prompt photon event selection~\cite{atlassusy}.
In the regions below and left of the lines a discovery could be made
with the corresponding amount of data assuming negligible background.
In the high $\tan\beta$ region the $\tilde{\tau}$ is the NLSP and no
photons occur in the decay chain.
In general, it is found that the discovery potential extends to
significant regions of the parameter space not excluded by the current
limits, even with early LHC data.
Similar studies have been performed for CMS with comparable results.

\begin{footnotesize}

\end{footnotesize}


\begin{thebibliography}{99}

\bibitem{d0wg} V.~M.~Abazov {\it et al.} (D\O\ Collaboration), Phys.\ Rev.\ Lett. {\bf 100} 241805 (2008).

\bibitem{wz} A.~Abulencia {\it et al.} (CDF Collaboration), Phys.\ Rev.\ Lett. {\bf 98} 161801 (2007);\\
D.~Acosta {\it et al.} (CDF Collaboration), Phys.\ Rev {\bf D71} 091105 (2005).

\bibitem{cdfzz} T.~Aaltonen {\it et al.} (CDF Collaboration), Phys.\ Rev.\ Lett. {\bf 100} 201801 (2008).

\bibitem{d0zz} V.~M.~Abazov {\it et al.} (D\O\ Collaboration), arXiv:0808.0703~[hep-ex] (2008).

\bibitem{wanke} R.~Wanke, arXiv:0807.2735~[hep-ex] (2008).

\bibitem{zeusCC} ZEUS Collaboration, contributed paper to {\it HEP-EPS 2007}, abstract {\bf 85}, ZEUS-prel-07-023.

\bibitem{cdfhelicity}  CDF Collaboration, CDF Note 9431 (2008);\\
CDF Collaboration, CDF Note 9114 (2008).

\bibitem{d0tb} V.~M.~Abazov {\it et al.} (D\O\ Collaboration), Phys.\ Rev.\ {\bf D78} 012005 (2008).

\bibitem{zeusEW} ZEUS Collaboration, contributed paper to {\it ICHEP 2006}, ZEUS-prel-06-003.

\bibitem{h1EW} H1 Collaboration, contributed paper to {\it HEP-EPS 2007}, abstract {\bf 230}, H1prelim-07-041.

\bibitem{nutev} K.~McFarland, these proceedings and references therein.

\bibitem{topmass} CDF and D\O\ Collaborations, arXiv:0808.1089~[hep-ex] (2008).

\bibitem{d0mtxsec} V.~M.~Abazov {\it et al.} (D\O\ Collaboration), Phys.\ Rev.\ Lett.\ {\bf 100} 192004 (2008).

\bibitem{ttxsecnlonll} M.~Cacciari {\it et al.}, arXiv:0804.2800~[hep-ph] (2008).

\bibitem{ttxsecnnll} S.~Moch and P.~Uwer, arXiv:0804.1476~[hep-ph] (2008).

\bibitem{newhiggs} CDF and D\O\ Collaborations, arXiv:0808.0534~[hep-ex] (2008).
 
\bibitem{h1general} H1 Collaboration, contributed paper to {\it HEP-EPS 2007}, abstract {\bf 199}, H1prelim-07-061.

\bibitem{cdfgeneral} T.~Aaltonen {\it et al.} (CDF Collaboration), Phys.\ Rev.\ {\bf D78} 112002 (2008);\\
C.~Henderson (for the CDF Collaboration), arXiv:0805.0742~[hep-ex] (2008).

\bibitem{multilep} F.~D.~Aaron {\it et al.} (H1 Collaboration), submitted to Phys.\ Lett.\ {\bf B}, arXiv:0806.3987~[hep-ex] (2008);\\
ZEUS Collaboration, preliminary results ZEUS-prel-07-022, ZEUS-prel-08-006. 

\bibitem{multilepcombined}  H1 and ZEUS Collaborations, contributed paper to {\it LP 2007}, H1prelim-07-166, ZEUS-prel-07-024.

\bibitem{isolep} S.~Chekanov {\it et al.} (ZEUS Collaboration), arXiv:0807.0589~[hep-ex] (2008);\\
H1 Collaboration, contributed paper to {\it HEP-EPS 2007}, abstract {\bf 228}, H1prelim-07-063;\\
H1 Collaboration, contributed paper to {\it HEP-EPS 2007}, abstract {\bf 227}, H1prelim-07-064.

\bibitem{isolepcombined} H1 and ZEUS Collaborations, contributed paper to {\it HEP-EPS 2007}, abstract {\bf 196}, H1prelim-07-162, ZEUS-prel-07-029.

\bibitem{h1top} H1 Collaboration, contributed paper to {\it HEP-EPS 2007}, abstract {\bf 776}, H1prelim-07-163.

\bibitem{cdffcnc} CDF Collaboration, submitted to Phys.\ Rev.\ Lett., arXiv:0805.2109~[hep-ex] (2008).

\bibitem{cdftprime} T.~Aaltonen {\it et al.} (CDF Collaboration), Phys.\ Rev.\ Lett.\ {\bf 100} 161803 (2008);\\
CDF Collaboration, CDF Note 9446 (2008).

\bibitem{d0ttprime} D\O\ Collaboration, D\O\ Note 5600-CONF (2008).

\bibitem{cdfpairs} CDF Collaboration, CDF Note 9246 (2008).

\bibitem{d0excited} V.~M.~Abazov {\it et al.} (D\O\ Collaboration), Phys.\ Rev.\ {\bf D77} 091102(R) (2008).

\bibitem{h1excitedelec} F.~D.~Aaron {\it et al.} (H1 Collaboration), submitted to Phys.\ Lett.\ {\bf B}, arXiv:0805.4530~[hep-ex] (2008).

\bibitem{h1excitednu} F.~D.~Aaron {\it et al.} (H1 Collaboration), Phys.\ Lett.\ {\bf B663} 382 (2008).

\bibitem{h1lq} H1 Collaboration, preliminary results, H1prelim-07-164.

\bibitem{h1lfv} H1 Collaboration, preliminary results, H1prelim-07-167.

\bibitem{beauchemin} P.~Beauchemin, these proceedings and references therein.

\bibitem{h1CI} H1 Collaboration, contributed paper to {\it LP 2007}, H1prelim-07-141.

\bibitem{zeusCI} ZEUS Collaboration, contributed paper to {\it ICHEP 2006}, ZEUS-prel-06-018.

\bibitem{krutelyov} CDF Collaboration, submitted to Phys.\ Rev.\ Lett., arXiv:0807.3132~[hep-ex] (2008).

\bibitem{squarksandgluons} V.~M.~Abazov {\it et al.} (D\O\ Collaboration), Phys.\ Lett.\ {\bf B660} 449 (2008);\\
CDF Collaboration, CDF Note 9229 (2008).

\bibitem{charginos} CDF Collaboration, submitted to Phys.\ Rev.\ Lett., arXiv:0808.2446~[hep-ex] (2008);\\
D\O\ Collaboration, D\O\ Note 5464-CONF (2007).

\bibitem{bruneliere} R.~Bruneli\`{e}re, these proceedings and references therein.

\bibitem{d0gmsb} V.~M.~Abazov {\it et al.} (D\O\ Collaboration), Phys.\ Lett.\ {\bf B659} 856 (2008).

\bibitem{atlassusy} ATLAS Collaboration, ATLAS Note SUSY CSC 8, to be published.

\end{thebibliography}
\end{document}